\documentstyle[aps,pra,twocolumn]{revtex}
\begin{document}
\draft
\twocolumn[\hsize\textwidth\columnwidth\hsize\csname @twocolumnfalse\endcsname
\title{Universality in Heavy Fermions Revisited}
\author{Mucio A.Continentino}
\address{Instituto de Fisica,
Universidade Federal Fluminense \protect\\
Campus da Praia Vermelha, Niter\'oi, 24.210-340, RJ, Brasil}
\date{\today}
\maketitle
\begin{abstract}
A previous scaling analysis of pressure experiments in heavy fermion  is
reviewed and enlarged.  We show that the critical exponents obtained from this
analysis indicate that a one-parameter scaling describes these experiments. We
obtain explicitly the
enhancemente factors showing that these systems are indeed near criticality and
that the  scaling approach is appropriate. The physics responsible for the
one-parameter scaling and breakdown of hyperscaling is clarified. We discuss a
microsocopic theory that is in agreement with the experiments.
The scaling theory is generalized for the case the {\em shift} and {\em
crossover} exponents are different. The exponents governing the physical
behavior along the non-Fermi liquid trajectory are obtained for this case.
\end{abstract}

\pacs{PACS Nos. 71.27+a 75.30Mb 71.10Hf 75.45+j 64.60Kw}
] \newpage

\section{Introduction}

The scaling theory of heavy fermions is based on the existence of a quantum
critical point (QCP) which governs the physical behavior of these systems
\cite{mucio1,mucio2,mucio3,mucio4}. It yields the following scaling
relations for the singular part of the free energy density
\begin{equation}
f\propto {|\delta |}^{2-\alpha }F\left[ \frac T{T_{coh}},\frac H{{|\delta |}%
^{\beta +\gamma }},\frac h{h_c}\right]
\end{equation}
where the {\em coherence temperature}, $T_{coh}={|\delta |}^{\nu z}$ and the
characteristic uniform field, $h_c={|\delta |}^{\phi _h}$. $H$ and $h$ are
staggered and uniform magnetic fields respectively. The scaling properties
of relevant thermodynamic quantities, like the uniform susceptibility, $\chi
_0\propto \partial ^2f/\partial h^2$, the thermal mass, $m_T = C/T \propto
\partial ^2f/\partial T^2$, for $T \ll T_{coh}$ are obtained from Eq.1. The
critical exponents obey standard scaling relations but hyperscaling is
modified due to the quantum character of the critical point. It is given by $%
2-\alpha =\nu (d+z)$ where $z$ is the dynamic exponent and $d$ the dimension
of the system \cite{mucio2}. The quantity $\delta =(J/W)-(J/W)_c$ measures
the distance to the QCP. $J$ and $W$ are parameters of the Kondo lattice
Hamiltonian, the coupling between localized and conduction electrons and the
bandwidth, respectively.

\section{Scaling Analysis}

In heavy fermions the ratio $(J/W)$ depends on volume $V$ and consequently
on pressure $P$. Let us define the critical volume $V_c$ as the volume at
which $(J/W)=(J/W)_c $. Consider a physical quantity which close to $V_c$
behaves as $X(P)=A{|(V-V_c)/V_c|}^{-x}$ where $V$ is the volume at pressure $%
P$. If we introduce a reference pressure $P_0$ (volume $V_0$) we have:
\begin{equation}
\frac{X(P)}{X(P_0)}={\ \left( \frac{V-V_c}{V_0-V_c}\right) }^{-x}={\ \left(
\frac{V-V_0+V_0-V_c}{V_0-V_c}\right) }^{-x}
\end{equation}
Defining, $\delta_0 =V_0-V_c$ and $\Delta V=V-V_0$, where the latter gives
the change in volume due to the pressure change $\Delta P=P-P_0$, we get
\begin{equation}
\frac{X(P)}{X(P_0)}={\left( \frac{\Delta V+\delta_0 }{\delta_0} \right) }%
^{-x}={\left( 1+\frac{\Delta V}{\delta_0} \right) }^{-x}
\end{equation}
Then
\begin{equation}
Ln\left[ \frac{X(P)}{X(P_0)}\right] =-xLn\left( 1+\frac{\Delta V}{\delta_0}
\right)
\end{equation}
For $\Delta V$ sufficiently small, or small changes of pressure from the
reference pressure, we have $\Delta V=-\kappa _0V_0\Delta P$ where the
compressibility is given by $\kappa _0= (-1/V)(\partial V / \partial P)$. In
the limit that $\left( \Delta V/\delta_0 \right) \ll 1$ we obtain
\begin{equation}
Ln\left[ \frac{X(P)}{X(P_0)}\right] \approx x\kappa _0(V_0/\delta_0 )\Delta P
\end{equation}
where $(V_0/\delta_0 )=V_0/(V_0-V_c)= \alpha _V /(\alpha _V - 1)$ with $%
\alpha _V=V_0/V_c$. The equation above holds if the system at the reference
pressure $P_0$ is not too close to the quantum critical point, otherwise the
condition $\left( \Delta V/\delta_0 \right) \ll 1$, or equivalently $(\kappa
_0 \Delta P \alpha _V)/(1-\alpha _V) \ll 1$ is not satisfied. Note that in
Eq.5 the coefficient of $\Delta P$ depends on the critical exponent $x$
associated with the physical quantity $X$. The validity of Eq. 5 for several
physical quantities at and below the coherence line has been verified for
the heavy fermions $CeRu_2Si_2$, $CeAl_3$, $UPt_3$ and $CeCu_6$ as shown in
Fig.1 \cite{mucio2}. These materials are located in the non-critical side ($%
V<V_c$) of Doniach's phase diagram. For $CeAl_3$ a reference pressure of $%
1.2 $ kbars has been used to guarantee that this is the case, otherwise $%
P_0=0$. The collapse of the different data for a given material on a single
line, as shown in Fig.1, implies the following relations among the critical
exponents \cite{mucio2}, $2-\alpha =\nu z$ and $\phi _h=\nu z$. The meaning
and implications of these relations will be discussed further down. The
inclination of the lines in Fig. 1, i.e., $\Gamma _V=Ln\left[ \frac{X(P)}{%
X(P_0)}\right] /\Delta P$, for different compounds are given in Table 1.
{}From Eq.5 we note that the Gr\"{u}neisen parameters $\Omega _V=(\Gamma
_V/\kappa _0)=x \alpha _V/(\alpha _V - 1)$ provide essentially the
enhancement factors due to the proximity of the quantum phase transition.
Assuming, for example, that $x=-1$ as we will discuss below, the results for
$\Omega _V$ in Table 1 yield values of $\alpha _V$ ranging from $\alpha
_V\approx 0.97$ to $\alpha _V\approx 0.99$ as shown in this Table. This
clearly indicates that the systems we are considering are close to the QCP
and a scaling analysis is justified.

Consider the case of $CeCu_6$. Eq.1 yields, $m_T = C/T \propto \partial
^2f/\partial T^2 \propto {\vert \delta \vert}^{2 - \alpha - 2\nu z}$. From
the experimental relation $2 - \alpha = \nu z$, we get ${m_T}^{-1} \propto {%
\vert (V_0 - V_c)/V_c \vert}^{\nu z}=(1 - \alpha_V)^{\nu z}$. Taking $x=-1$,
i.e., $\nu z=1$, as before (see also below), we get an enhancement factor $%
\Omega _V/\alpha_V = 1/(1-\alpha _V)$ for the thermal mass, $m_T$, of this
material of approximately $120$.

The data in Table 1 calls attention to the fact that $\Omega _V$ is larger
for $CeRu_2Si_2$, although this system has the smallest thermal mass of the
four compounds \cite{thompson}. In order to conciliate this result with the
idea of universality, i.e., that the critical exponents are the same for a
given quantity independent of the material, it is important to write the
enhancement factor in terms of the parameters of the Kondo lattice
Hamiltonian. For this purpose we need a relation between $(J/W)$ and the
volume. We assume that $j(V)=(J/W)=(J/W)_0exp(-q(\frac{V-V_0}{V_0}))$ where $%
(J/W)_0$ is the value of this ratio at the reference pressure $P_0$ or
volume $V_0$ and $q$ a material-dependent parameter \cite{lavagna}. For
sufficiently small volume changes we get $q(1-\alpha _V)/\alpha _V=(1-\alpha
_J)$ where $\alpha _J=\frac{(J/W)_c}{(J/W)_0}$ and $(J/W)_c=j(V_c)$. This
allows us to write
\begin{equation}
\Omega _V=\frac{\Gamma _V}{\kappa _0}=\frac{-xq}{1-\alpha _J}
\end{equation}
where the parameter $q$, which relates changes in volume to changes in the
interactions, depends on the particular system. Taking $q=5$, for $%
CeRu_2Si_2 $ \cite{lavagna} and $\nu z=1$, as before, we get an enhancement
for the thermal mass of approximately $36$ for this system. In this way
taking different values of $q$ for different systems we may explain the
hierarchy of thermal masses using the same critical exponents. Note that for
$CeAl_3$ the reference pressure $P_0=1.2$kbars such that $m_T$ for this
pressure is already very much reduced.

Let us return now to discuss the scaling relations obtained from the data of
Figure 1, namely, $2-\alpha =\nu z$ and $\phi _h=\nu z$. It is easy to
verify that these equations imply a simple one-parameter scaling such that $%
\chi _0^{-1} \propto m_T^{-1} \propto h_c \propto A^{-1/2} \propto T_{coh} $%
, where $A$ is the coefficient of the $T^2$ term of the resistivity. As
concerns the relations among the thermodynamic quantities they arise from a
free energy obeying the simple scaling form, $f \propto {\vert \delta \vert}%
^{\nu z}F[ T/{\vert \delta \vert}^{\nu z}, h/{\vert \delta \vert}^{\nu z}]$.
This one-parameter scaling is reminiscent of single impurity and other
phenomenological, non-critical, approaches to the heavy fermion problem \cite
{thompson}. In the former case the characteristic temperature is identified
with the Kondo temperature. There is however a fundamental difference
between these approaches and the present scaling theory. Here the
characteristic energy scale given by the {\em coherence temperature} $%
T_{coh}\propto {|(J/W)-(J/W)_c|}^{\nu z}$ vanishes at the critical point of
the Kondo lattice. Furthermore the scaling behavior found in these heavy
fermion systems is due in our approach to their proximity to the QCP at $%
(J/W)=(J/W)_c, T=0, H=0, h=0$. The vanishing of $T_{coh}$ at the QCP led to
the prediction \cite{pre} and observation of non-Fermi liquid behavior in
heavy fermion systems \cite{lohneysen}.

The observation of one-parameter scaling in a three dimensional ($d=3$)
critical theory with three independent exponents is clearly associated here
with the breakdown of the hyperscaling relation $2 - \alpha = \nu (d+z)$.
Note that the relation $2-\alpha = \nu z$ arising from the experimental data
is the hyperscaling relation for $d=0$. It is not surprising then, that it
yields scaling properties which are formally similar to those of a single
impurity problem. Violation of hyperscaling is not uncommon in critical
phenomena \cite{mucio3} and below we shall discuss the possible reasons it
occurs here.

One-parameter scaling can also result from the constraints imposed by the
Fermi liquid (FL) behavior below the coherence temperature. Let us assume
that the entropy of the local moments, of total angular momentum $J$, for $%
T>>T_{coh}$, $S(T>>T_{coh})=Nk_BLn(2J+1)$ goes into that of the Fermi liquid
which develops below $T_{coh}$ and is given by, $S(T \ll
T_{coh})=\int_0^{T_{coh}}dTC(T)/T$ with $C(T)=m_TT$. Equating both entropies
we find $m_T^{-1}\propto T_{coh}$, which in turn implies $2-\alpha =\nu z$.
This argument \cite{fisk} relies on the existence of a single characteristic
temperature in the non-critical side of the phase diagram which is not
necessarily always the case, as will be seen below \cite{mucio5}.

\section{Microscopic Model}

We now discuss a model of nearly antiferromagnetic Fer\-mi liquid \cite
{lacroix} which describes the results of the pressure experiments analyzed
above. We start with the expression for the singular part of the free energy
density of a nearly antiferromagnetic Fermi liquid due to spin fluctuations
\cite{moriya} \cite{hasegawa} \cite{taki}:
\begin{equation}
f=-\frac 1\pi \sum_q\int_0^\infty d\omega \coth (\beta \omega /2)\tan
^{-1}\left[ \frac{I\Im m\chi ^0(q,\omega )}{1-I\Re e\chi ^0(q,\omega )}%
\right]
\end{equation}
where
\begin{equation}
\chi ^0(Q+q,\omega )=\chi _Q(1-aq^2-b\omega ^2+ic\omega )
\end{equation}
for $q \ll Q$ and $\omega \ll 1$. At zero temperature we get
\begin{equation}
f=-\frac 1\pi \sum_q\int_0^\infty d\omega \tan ^{-1}\left[ \frac{I\chi
_Qc\omega }{1-I\chi _Q(1-aq^2-b\omega ^2)}\right]
\end{equation}
This can be rewritten as
\begin{equation}
f=-\frac 1\pi \int d\vec{q}\int d\omega \tan ^{-1}\left[ \frac{\omega \xi ^z%
}{1+q^2\xi ^2+|\delta |(\omega \xi ^z)^2}\right]
\end{equation}
where the correlation length $\xi ={|\delta |}^{-\nu }$ and the distance to
the QCP, $\delta =1-I\chi _Q$. The correlation length exponent assumes the
mean-field (or Gaussian) value, $\nu =1/2$ and the dynamic exponent, $z=2$,
typical of antiferromagnetic spin fluctuations. All constants have been
taken equal to $1$. In the limit $\delta \rightarrow 0$, a change of
variable yields the scaling properties of this model. We find $f\propto {%
|\delta |}^{\nu (d+z)}$, where $d$ is the dimension of the system, with
additional scaling corrections due to the term $|\delta |(\omega \xi ^z)^2$.
{}From Eq.1, with $F[0,0,0]=constant$ we obtain the generalized hyperscaling
relation, $2-\alpha =\nu (d+z)$ \cite{mucio1}.

Let us consider the case of localized spin fluctuations \cite{lacroix},
i.e., we neglect the q-dependence of $\chi ^0(q,\omega )$. We get
\begin{equation}
f\propto \frac{1}{\pi} \int d\omega \tan ^{-1}\left[ \frac{\omega \xi ^z}{%
1+|\delta |(\omega \xi ^z)^2}\right]
\end{equation}
Now for $\delta \rightarrow 0$, we find $f\propto {|\delta |}^{\nu z}$, with
scaling corrections. In this case, from Eq.1, we obtain $2-\alpha =\nu z$
and hyperscaling is violated due to the neglect of the q-dependence of $\chi
^0$. Note that in this local theory the exponent $\nu z =1$ as for the
nearly antiferromagnetic Fermi liquid.

For $T \ne 0$ the relevant expression of the singular part of the free
energy density is given by
\[
f \propto -T \int^{q_c}d\vec{q} \int^{x_c}dx \coth x \tan^{-1} [ xT \xi^z
(1-
\]
\begin{equation}
\frac{q^2 \xi^2}{1+q^2{\xi}^2} ) ]
\end{equation}
which can be written in the form
\begin{equation}
f \propto {\vert \delta \vert}^{\nu(d+z)}F[T/T_{coh}]
\end{equation}
for $x_c = \beta \omega_c/2 =1/2$ and $q_c \xi$ very large, as close to the
critical point.

Again neglecting the q-dependent contribution we obtain, from Eq.12, the
thermodynamic properties of the localized model. For $xT \xi^z \ll 1$, i.e.,
$\omega_c {\xi}^z \ll 1$ we may write $\tan^{-1}y \approx y$. Furthermore we
take $\hbar \omega_c = k_B T$, in which case the constraint $\omega_c {\xi}%
^z \ll 1$ implies $T \ll T_{coh} \propto {\vert \delta \vert}^{\nu z}$. In
this regime the free energy exhibits Fermi liquid behavior and is given by, $%
f \propto -(4/3) \pi q_c^3 g(1/2) T^2 \xi^z$, where $g(y) = \int_0^y dx x
\coth x $. The breakdown of the scale invariant form, Eq.13 and consequently
of hyperscaling is due to the neglect of the q-dependence. The thermal mass
is given by
\begin{equation}
m_T^L = - \partial^2 f / \partial T^2 = m_M^0 (1/3) q_c^2 \xi^z \propto {%
\vert \delta \vert}^{- \nu z} = {\vert \delta \vert}^{-1}
\end{equation}
and diverges at the transition. $m_M^0$ is a non-critical constant, which
depends on local parameters and, linearly, on $q_c$ \cite{moriya}.

Within the same Fermi liquid regime, for $T \ll T_{coh}$, the free energy
given by Eq.12 can be calculated and the thermal mass is given by, $m_T^M =
- \partial^2 f / \partial T^2 = m_M^0 \left[ 1- (q_c \xi)^{-1} tan^{-1}(q_c
\xi) \right]$. In this case the thermal mass increases but does not diverge
as the system approaches the critical point \cite{moriya}. The second term
in this expression for $m_T^M$ is the universal scaling contribution, i.e.,
independent of the cut-off for $q_c \xi \rightarrow \infty$. It is
proportional to $\xi^{-1} \propto {\vert \delta \vert}^{1/2}$ since $m_M^0
\propto q_c$.

The free energies in the Fermi liquid regime ($T \ll T_{coh}$) as a function
of the distance to the QCP, for the localized and nearly antiferromagnetic
(NAF) models are shown in Fig.2. Notice that for $q_c \xi < 1/2$, they
nearly coincide. This can be seen directly, expanding $m_T^M$ to obtain, $%
m_T^M (q_c \xi \ll 1) \approx m_M^0 (1/3) q_c^2 \xi^2 = m_T^L$ (see Eq.14).
Then, in this regime \cite{note1} {\em the scaling properties of both models
are the same} and they yield similar results, for example, for the relative
pressure variation of the thermal mass, $m_T(P_0)/m_T(P)$.

For the resistivity in the local approximation we find \cite{lederer}
\begin{equation}
\rho = \frac{\rho_0}{T \xi^z} \int_0^{\infty} dx \frac{x^2}{(e^{\frac{x}{T
\xi^z}}-1)(1 - e^{-\frac{x}{T \xi^z}})(1+x^2)}
\end{equation}
For $T \xi^z \ll 1$, i.e., $T \ll T_{coh} = \xi^{-z} = {\vert \delta \vert}%
^{\nu z}$, we get, $\rho (T \ll T_{coh}) = \rho_0 ({\pi^2 / 3}) {\left(
T/T_{coh} \right)}^2$ and for higher temperatures $\rho(T >> T_{coh}) =
\rho_0 ({\pi/2}) \left( T/T_{coh} \right)$ \cite{lederer}.

As concerns the magnetic field the experimental relation $\phi_h = \nu z$
implies that the characteristic field $h_c \propto {\vert \delta \vert}^{\nu
z}$. In fact, in the local approach, the uniform magnetic field $h$ simply
adds to the frequency a precession term \cite{sachdev}. Then at $T=0$
\begin{equation}
f\propto \frac{1}{\pi} \int d\omega \tan ^{-1}\left[ ( \omega + h) \xi ^z
\right]
\end{equation}
and $\chi_0(T=0) = \partial^2 f/\partial h^2 \propto |\delta|^{-\nu z}$ with
$\nu z =1$. This yields a diverging or enhanced uniform susceptibility but
this result ceases to be valid sufficiently close to the critical point ($%
q_c \xi > 1$).

It is clear from the results above that the theory of antiferromagnetic
local spin fluctuations, with a single characteristic energy scale, $T_{coh}
\propto {\vert \delta \vert}^{\nu z}$, such that $f(T=0) \propto {\vert
\delta \vert}^{\nu z} = T_{coh}$, $m_T \propto {\vert \delta \vert}^{- \nu
z} = T_{coh}^{-1}$, $\chi_0(T=0) \propto |\delta|^{-\nu z} = T_{coh}^{-1}$
and $\rho (T \ll T_{coh}) = AT^2$ with $A \propto T_{coh}^{-2}$ correctly
describes the pressure experiments we analyzed before. Furthermore in this
approach the crossover exponent $\nu z =1$. This provided the motivation for
assuming this value for $\nu z$ in the calculation of the enhancement
factors in Section II of this paper.

The theory of localized antiferromagnetic paramagnons can be summarized in
the scaling form of the free energy, $f \propto |\delta|^{\nu
z}F[T/|\delta|^{\nu z},h/|\delta|^{\nu z}]$ with $\nu z=1$. Although this is
in agreement with the pressure experiments in a region of the phase diagram,
for $(J/W)>(J/W)_c$, $T \le T_{coh}$, as the systems get closer to the
critical point and $q_c \xi \rightarrow \infty$ the full $q$-dependence of
the dynamic susceptibility must be taken into account. The local theory also
does not describe the non-Fermi liquid behavior observed at $(J/W)=(J/W)_c$,
$T \rightarrow 0$. For example, it's result for the specific heat at $%
|\delta |=0$ is given by, $C = (3/2) Nk_B$, i.e., that of a classical gas of
N free particles. Sufficiently close to the $QCP$ it is then necessary to
consider the $q$-dependence of $\chi^0(Q+q,\omega )$. Taking this into
account we obtain the universal contribution for the free energy at the $QCP$%
, $f[\delta=0] \propto T^{\frac{d+z}{z}}$, in agreement with Eq.13, and the
specific heat $C/T(\delta=0) \propto \partial^2 f/\partial T^2 \propto
T^{1/2}$ for $d=3$, $z=2$ with $\nu z = 1$ \cite{hasegawa,millis,taki}.

\section{Generalized Scaling}

It is important to point out that the result $C/T\propto T^{1/2}$ at $%
|\delta |=0$ which arises from the scaling form Eq.13 is valid only in the
case of {\em extended scaling}, i.e., $\psi =\nu z$. Here $\psi $ is the
shift exponent such that the Neel temperature, $T_N\propto {|\delta |}^\psi $%
, close to the QCP \cite{mucio5} (see Fig.3). It turns out, experimentally,
that $\psi =1$ \cite{lohneysen} however, theoretically, one gets, $\psi
=z/(d+z-2)=2/3\ne \nu z=1$ \cite{hasegawa,millis,taki}. In this case a more
general scaling form for the free energy is required to describe the complex
critical behavior in the neighborhood of the QCP \cite{mucio5}. It is given
by \cite{mucio5},
\[
f\propto {|\delta (T)|}^{2-\alpha }F_c[t]
\]
\begin{equation}
t=\frac T{{|\delta (T)|}^{\nu z}}
\end{equation}
with
\[
\delta (T)=\delta (T=0)-T^{1/\psi }
\]
The singularities along the Neel line, ${|\delta (T)|}=0$, are described by
a new set of {\em tilde} exponents $\tilde{\alpha}$, $\tilde{\nu}$, etc.,
different from those associated with the zero temperature fixed point (the
{\em non-tilde} exponents). The scaling function $F_c[t=0]=$ $constant$ and $%
F_c[t\rightarrow \infty ]\propto t^x$ with $x=(\tilde{\alpha}-\alpha )/\nu z$
such that close to the critical Neel line we obtain the correct asymptotic
behavior, $f\propto A(T){|\delta (T)|}^{2-\tilde{\alpha}}$, where the
amplitude $A(T)=T^{\frac{\tilde{\alpha}-\alpha }{\nu z}}$. It is easy to
verify that in the case of extended scaling, i.e., $\psi =\nu z$, the
exponents associated with the zero temperature fixed point are sufficient to
describe the behavior along the non-Fermi liquid trajectory, $|\delta |=0$, $%
T\rightarrow 0$ and one finds the previous result, $C/T\propto T^{\frac{d-z}z%
}$. However in the situation here, where $\psi =2/3<\nu z=1$, the {\em tilde}
exponents also play a role in determining the behavior along this path,
which is tangent to the Neel line at the quantum multicritical point as
shown in Fig.3. We find for the specific heat
\begin{equation}
C/T\propto T^{\frac{(2-\tilde{\alpha})(\nu z-\psi )+\nu \psi (d-z)}{\nu
z\psi }}
\end{equation}
Assuming thermal Gaussian exponents, essentially $\tilde{\alpha}=1/2$, we
get, $C/T\propto T^{5/4}$ for $\psi =2/3$, $\nu =1/2$ and $z=2$, instead of $%
C/T\propto T^{1/2}$ for the case of extended scaling. The staggered
susceptibility $\chi _Q(\delta =0,T)\propto T^{-\tilde{\gamma}/\psi
}=T^{-3/2}$ since $\gamma =\tilde{\gamma}=1$ \cite{moriya}.

An interesting possibility has been raised by R\"osch at al. \cite{rosch}
who claimed that two-dimensional fluctuations are those relevant to describe
the observed critical behavior. In this case scaling is extended since $\psi
= z/(d+z-2) = \nu z= 1$. This approach leads to $C/T(\delta=0) \propto LnT$
at the QCP and in the Fermi liquid regime, for $T \ll T_{coh}$, to $m_T
\propto Ln|\delta|$.

\section{Above the Upper Critical Dimension}

In quantum phase transitions the relevant dimensionality is $d_{eff} = d +z$%
, as is evident from the modified hyperscaling relation, $2 - \alpha = \nu
(d +z)$. In the problems we have studied above, it turns out that $d_{eff} >
d_c$, where, $d_c = 4$ is the upper critical dimension for these magnetic
transitions. This is the reason Gaussian theories, as the SCR theory of spin
fluctuations \cite{moriya}, provide an adequate description of the quantum
critical point in nearly ferro and antiferromagnetic $3d$ materials \cite
{millis}. Let us consider further implications of the fact that $d+z > d_c =
4$. Consider the expression for the singular part of the $T=0$ free energy, $%
f \propto {\vert \delta \vert}^{\nu(d+z)}$. Since the Gaussian exponent $\nu
=1/2$, whenever $d+z > 4$, we can rewrite it as, $f \propto {\vert \delta
\vert}^{2 - \alpha}$ with $\alpha < 0$. In this case an analytic expansion
of the free energy close to the critical point, such that, $f \propto {\vert
\delta \vert}^{2}$, will always dominate the Gaussian contribution for $%
\delta$ sufficiently small. The total free energy, in the non-critical side
of the phase diagram, below the coherence line, can be written as a sum of
an analytic term and a Gaussian contribution,
\begin{equation}
f_t = {\vert \delta \vert}^{2} + {\vert \delta \vert}^{\nu(d+z)}F[\frac{T}{{%
\vert \delta \vert}^{\nu z}}]
\end{equation}
where we neglected the temperature dependence of the analytic part assuming
it is less singular than that of the Gaussian. In the Fermi liquid regime,
i.e., $T \ll T_{coh} ={\vert \delta \vert}^{\nu z}$, we have \cite{mucio2},
\begin{equation}
f_t = {\vert \delta \vert}^{2} + {\vert \delta \vert}^{\nu(d+z)} \left\{ 1 +
{\ \left( \frac{T}{{\vert \delta \vert}^{\nu z}} \right) }^2 + \cdots
\right\}
\end{equation}
then, sufficiently close to the critical point and for $d+z >4$, with $\nu
=1/2$, we obtain,
\begin{equation}
f_t = {\vert \delta \vert}^{2} + {\vert \delta \vert}^{\nu(d+z)} {\ \left(
\frac{T}{{\vert \delta \vert}^{\nu z}} \right) }^2 + \cdots
\end{equation}
which can be rewritten in the scaling form
\begin{equation}
f_t = {\vert \delta \vert}^{2}F[\frac{T}{T_{sf}}]
\end{equation}
where the value of $\alpha=0$ in the equation above, see Eq.1, can be
associated with the breakdown of hyperscaling for $d+z >4$. The new
spin-fluctuation temperature is given by
\begin{equation}
T_{sf} = {\vert \delta \vert}^{1 - \frac{d-z}{4}}
\end{equation}
where we used $\nu = 1/2$. Consequently the effect of the analytic
contribution is to introduce a new energy scale in the Fermi liquid region
of the phase diagram \cite{beal}, namely $T_{sf}$. If we calculate the
specific heat from the above expression for the free energy, we get, $C/T
\propto {\vert \delta \vert}^{\nu(d-z)}$ which of course coincides with the
Gaussian result. On the other hand taking into account the field dependence
of the analytic (mean-field) and Gaussian contributions, it can be easily
shown that the order parameter linear susceptibility, $\chi_0 = \left(
\partial^2 f/\partial h^2 \right)_{h=0}$ with $f \propto {\vert \delta \vert}%
^{2}F_0(T/T_{sf}, h/{\vert \delta \vert}^{\beta + \gamma})$ is given by $%
\chi_0 = {\vert \delta \vert}^{-1}F_1(T/T_{sf})$, for $T \ll T_{coh}$, where
we used the mean field exponents, $\beta=1/2$ and $\gamma=1$. The field $h$
here is that conjugated to the order parameter.

We point out also that, because $d+z >4$, even at ${\vert \delta \vert}=0$
there is a characteristic field $h_{cross}=J$ which yields the crossover
from mean-field to Gaussian behavior \cite{mucio5}. $J$ is the actual
critical coupling between localized and itinerant electrons \cite{mucio3}.
The order parameter $m$ at the quantum critical point varies with the
conjugated field $h$, according to $m \propto (h/J)^{1/\delta}$ \cite{mucio3}%
. For small fields, i.e., $h \ll h_{cross}$, the mean-field contribution
with $\delta_{MF} =3$ dominates while in the opposite case the Gaussian one
with $\delta_G = \frac{d+z+2}{d+z-2}$ \cite{mucio5} is dominant (in the
situation of interest here $d=3$ and $z=2$ such that $\delta_G = 7/3 <
\delta_{MF} =3$).

Finally note that the above discussion does not affect the local spin
fluctuation results since, in this case the $T=0$ free energy is more {\em %
singular} than the analytic contribution, at least for $\nu z < 2$.

\section{Conclusions}

Our analysis of the pressure dependence of several physical quantities for
different heavy fermions on a region of the phase diagram, for $%
(J/W)>(J/W)_c $, $T \le T_{coh}$, has shown that these systems are close to
a quantum critical point and a scaling approach is indeed appropriate. We
have obtained from the experimental data, Gr\"{u}neisen parameters which
yield the enhancement factors due to the proximity of the QCP. It turns out
that a one-parameter scaling is sufficient to describe the experiments in
this region of the phase diagram and this is associated with the violation
of the hyperscaling relation. This one-parameter scaling here is different
from that of single impurity approaches where the characteristic energy
scale is identified with the Kondo temperature. In our case $T_{coh}$
vanishes at the quantum critical point and this leads to the appearance of
non-Fermi liquid behavior at $\delta =0$. We have shown that a theory of
localized spin fluctuations, with $\nu z = 1$, describes the pressure
experiments summarized in Fig.1. {\em For these experiments the $q$%
-dependence of the dynamic susceptibility plays no role}. Physically this
must be related to the fact that the actual spectrum of spin fluctuations is
nearly $q$-independent, at least in the relevant directions of $q$-space, in
this part of the phase diagram. This region corresponds to the case $q_c \xi
< 1/2$ of the NAF model since, as we have shown, the scaling properties of
this and the local model are the same in this regime. In fact the local
approach can be obtained as an expansion of the full $q$-dependent theory of
antiferromagnetic spin fluctuations for $q_c \xi \ll 1$ (see also \cite
{note1}). Then it holds in a region, {\em not too close to the QCP}, but
where scaling, specific to this q-independent regime (one-parameter
scaling), still applies. The local theory allows for an explicit calculation
of the Wilson $(m_T/\chi_0)$ and Kadowaki-Woods $(A/m^2_T)$ ratios \cite{pre}
which turn out to be constants, i.e., independent of the distance to the
critical point.

As the system moves closer to the critical point ($q_c\xi \rightarrow \infty
$) we leave the limit of validity of the local theory and the full
q-dependent susceptibility must be used. In both cases $\nu z=1$. In the
q-dependent regime a generalized scaling theory is required to describe the
complex critical behavior in the neighborhood of the QCP since the {\em shift%
} and crossover exponents are different for nearly antiferromagnetic $3d$
systems. This is particularly relevant along the {\em non-Fermi liquid
trajectory} which is tangent to the $QCP$ for $\psi =2/3<1$. It turns out
that the exponents along this line depend on the thermal exponents
characterizing the singularities on the Neel line, besides those associated
with the $QCP$.

Although for the systems investigated the effective dimension $d+z > d_c$,
it is the local nature of the spin fluctuation spectrum in the part of the
phase diagram investigated that is responsible for the breakdown of
hyperscaling and one-parameter scaling, observed in the experiments. In the
microscopic approach the violation of hyperscaling can be traced to the
cut-off acting as a relevant variable.

The purpose of a scaling theory of heavy fermions is to provide a unified
description, in terms of a set of critical exponents, not only of the
non-Fermi liquid regime but also of both sides of the quantum critical
point. While most of the studies have been carried on in the non-critical
side we also expect to find scaling behavior on the ordered region of the
phase diagram, i.e., for $(J/W) < (J/W)_c$, where the systems at $T=0$ have
long range magnetic order. We hope future studies will also include this
interesting region and also a larger class of materials \cite{steglich}.

\acknowledgements{I would like to thank Sergey Budko and Claudine Lacroix
for discussions and a critical reading of the manuscript. Also I thank the
LEPES-CNRS and in particular J-L. Tholence for the hospitality in Grenoble,
where part of this work was carried out.}

\begin{center}
{\bf APPENDIX: }
\end{center}

We discuss here the limit $q_c\xi \ll 1$ in relation to the theory of Ref.
\cite{taki}. The free energy is given by,
\begin{equation}
f=-\frac{3T}{\pi} \int_0^{q_c}d\vec{q}\int_0^{x_c} dx \coth x \tan
^{-1}\left[ \frac{x T }{J_Q - J_Q^c + Aq^2} \right]
\end{equation}
where $x=\frac{\beta \omega}{2}$. $J_Q$ is the q-dependent exchange coupling
between the local moments at the wavevector $Q$ of the incipient magnetic
instability and $J_Q^c$ its critical value. $A$ is the {\em stiffness} of
the spin fluctuations defined by the small wavevector expansion of the
magnetic coupling close to the wavevector $Q$, i.e., $J_Q-J_{Q+q}=Aq^2+%
\cdots $.

A redefinition of the relevant quantities yields,
\begin{equation}
f=-\frac{3T}\pi \int_0^{q_c}d\vec{q}\int_0^{x_c}dx\coth x\tan ^{-1}\left[
\frac{xT\xi ^z}{A(1+q^2\xi ^2)}\right]
\end{equation}
where the correlation length $\xi $ may be written as, $\xi =\alpha \xi _L$
with $\alpha =(A/J_Qa^2)^{1/2}$, $\xi _L=a|\delta |^{-1/2}$ and $\delta
=1-(J_Q^c/J_Q)$, such that, $\nu =1/2$. $a$ is an interatomic distance and
the dynamic exponent $z=2$. Since $1/(1+y^2)\le 1$, for $\frac{x_cT\xi ^z}A%
\ll 1$, or equivalently, $\omega _c\xi _L^z\ll 1$, which assuming $\hbar
\omega _c=k_BT$ implies $T\ll T_{coh}\propto \xi _L^{-z}$, we can expand the
$\tan ^{-1}$ for small arguments. In this regime the free energy exhibits
Fermi liquid behavior. % and is given by
%\begin{equation}
%f \approx  -\frac 3\pi \frac{T^2 \xi^z}{A} g(1/2)
%\int_0^{q_c}   dq  \frac{4 \pi q^2   }{ 1  + q^2 \xi^2}
%\end{equation}
%with $g(y)= \int_0^{y} x \coth x dx$. Integrating we get,
Performing the integrations, we get
\begin{equation}
f\approx -\frac{12T^2\xi ^{(z-d)}}Ag(1/2)q_c\xi \left( 1-\frac{\tan
^{-1}q_c\xi }{q_c\xi }\right)
\end{equation}
where $g(y)=\int_0^yx\coth xdx$. The limit $q_c\xi \ll 1$ is obtained,
either because the system is not too close to the QCP, i.e., $|\delta |$ is
large, or $A\rightarrow 0$ due to the localization of the fluctuations, we
get
\begin{equation}
f\propto \frac{12T^2}Ag(1/2)q_c\left[ \frac 13(q_c\xi )^2-\frac 15(q_c\xi
)^4+\cdots \right]
\end{equation}
Using $\xi =\alpha \xi _L$, as defined above, we get
\begin{equation}
f\approx -\frac{4T^2\xi _L^2}{J_Qa^2}g(1/2)q_c^3+\frac{12AT^2\xi _L^4}{%
5J_Q^2a^4}g(1/2)q_c^5+O(A^2)
\end{equation}
The first term is independent of $A$ and corresponds to the localized model
where the stiffness $A$ vanishes. The validity of the local theory is then
given by $q_c\xi \ll 1$ or $q_c\xi _L\ll (J_Qa^2/A)^{1/2}$ such that when $%
A\rightarrow 0$ the local model becomes valid arbitrarily close to the QCP.

%\end{thebibliography}

\newpage
%\vspace{1cm}

\begin{center}
{\large TABLE }
\end{center}

\begin{table}[h]
\begin{center}
\par
\begin{tabular}{|c|c|c|c|c|}
\hline
Compound & $\Gamma_V$ & $\kappa_0$ & $\Omega_V$ = $\Gamma_V$/$\kappa_0$ & $%
\alpha_V = V_0/V_c$ \\ \hline
$CeAl_3$ & 89 & 2.17 & 41 & 0.976 \\ \hline
$CeRu_2Si_2$ & 171 & 0.95 & 180 & 0.994 \\ \hline
$UPt_3$ & 26 & 0.48 & 54 & 0.981 \\ \hline
$CeCu_6$ & 133 & 1.1 & 121 & 0.992 \\ \hline
\end{tabular}
\par
\end{center}
\caption{Gr\"uneisen parameters for different heavy fermions according to
the relative pressure variation of several physical quantities shown in Fig.
1 \protect\cite{mucio2}. $\Gamma_V$ and the compressibility $\kappa_0$ are
in $Mbar^{-1}$. The reference pressure for $CeAl_3$ is $P_0=1.2$ kbars
otherwise $P_0=0$. The data for $CeCu_6$ is taken from Ref. \protect\cite
{thompson} and references therein.}
\end{table}

%newpage

\begin{center}
{\large FIGURE CAPTIONS }
\end{center}

Figure 1. Semi-logarithmic plot of $X(P)/X(P_0)$ for several physical
quantities, at or below $T_{coh}$, as a function of pressure for different
heavy fermions \cite{mucio2}. For $CeAl_3$, $P_0=1.2$ kbars otherwise $P_0=0$%
. The numbers close to the lines are their inclinations, $\Gamma_V$, given
in Table I (also see text).

\vspace{1cm}

Figure 2. The free energy in the Fermi liquid regime, at a fixed temperature
$T_0 \ll T_{coh}$, for the localized and nearly antiferromagnetic models as
a function of the distance to the critical point, located at $(q_c
\xi)^{-1}=0$. For $(q_c \xi)^{-1} > 2$ the free energies of both models
nearly coincide and consequently have the same scaling behavior.

\vspace{1cm}

Figure 3. Phase diagram of a nearly antiferromagnetic $3d$ system, with the
shift exponent $\psi =2/3$ and the crossover exponent $\nu z =1 $. The {\em %
tilde} (thermal) exponents determine the critical behavior on the Neel line.
In this case of $\psi <1$, the non-Fermi liquid trajectory is tangent to the
quantum critical point and the thermal exponents besides those associated
with the $QCP$ are required to characterize the physical behavior along this
line ($|\delta|=0, T \rightarrow 0$).


\begin{references}
\bibitem{note1}  It is interesting to discuss the limit $q_c\xi \ll 1$ in
relation to the theory of Ref. \cite{taki}, which is formally similar to the
one presented here. In this approach the correlation length as appears in
Eqs. 10 and 12 is given by, $\xi =(A/|\delta |)^{1/2}$, where $A$ is the
{\em stiffness} of the spin fluctuations. It is defined by $%
J_Q-J_{Q+q}=Aq^2+\cdots $ with $J_Q$ the q-dependent exchange coupling at
the wavevector $Q$ of the incipient instability. Then, in this approach, the
limit $q_c\xi \ll 1$ is obtained either because the system is not too close
to the QCP, i.e., $|\delta |$ is large, or $A$ is small due to the
localization of the fluctuations. In this limit one recovers again from this
theory the thermodynamic properties of the local model.
\end{references}
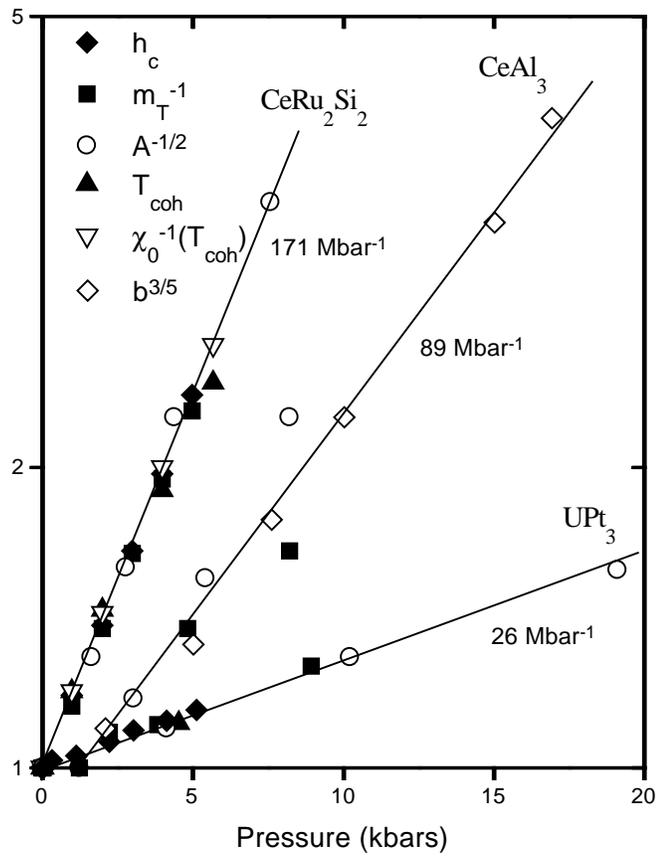
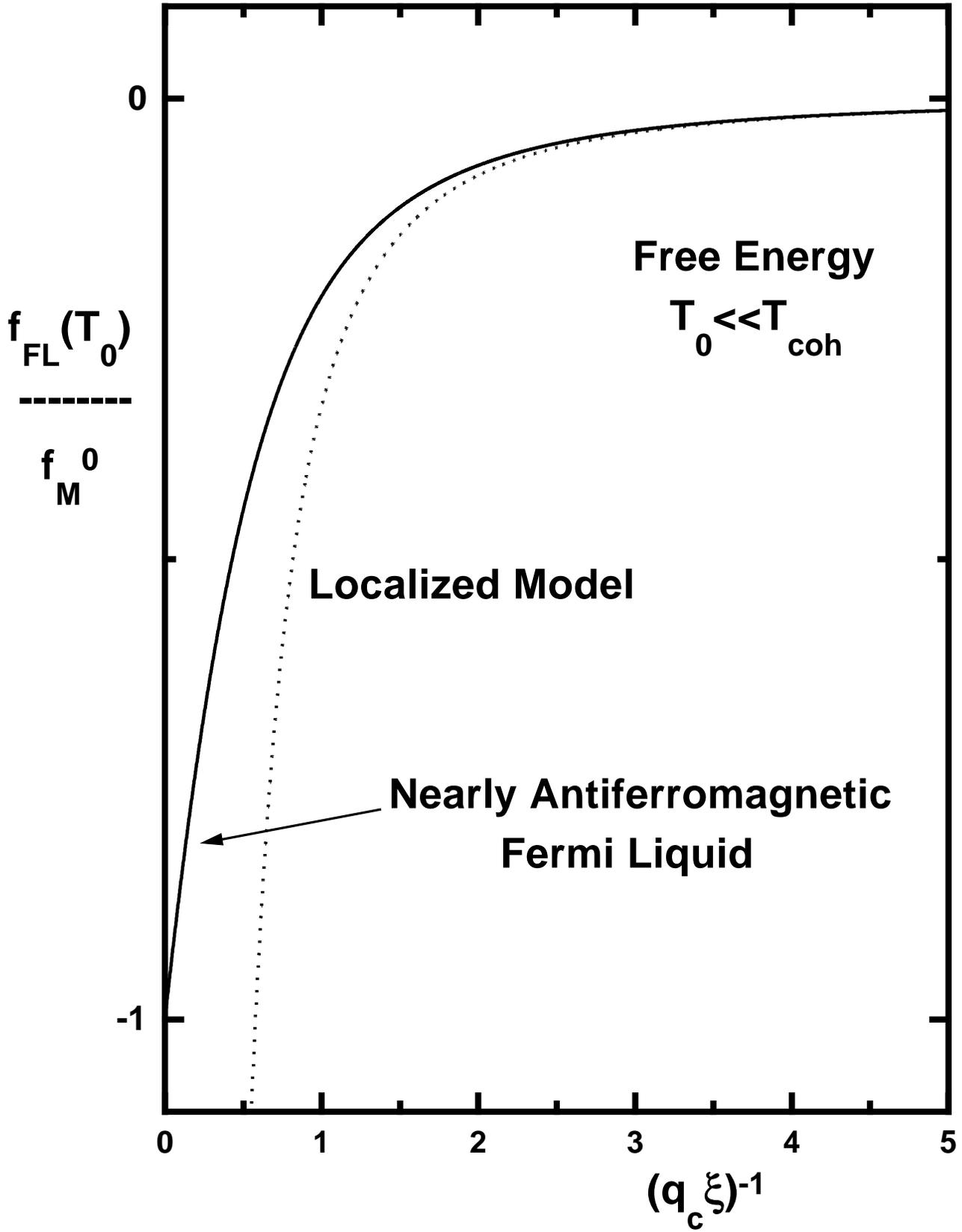
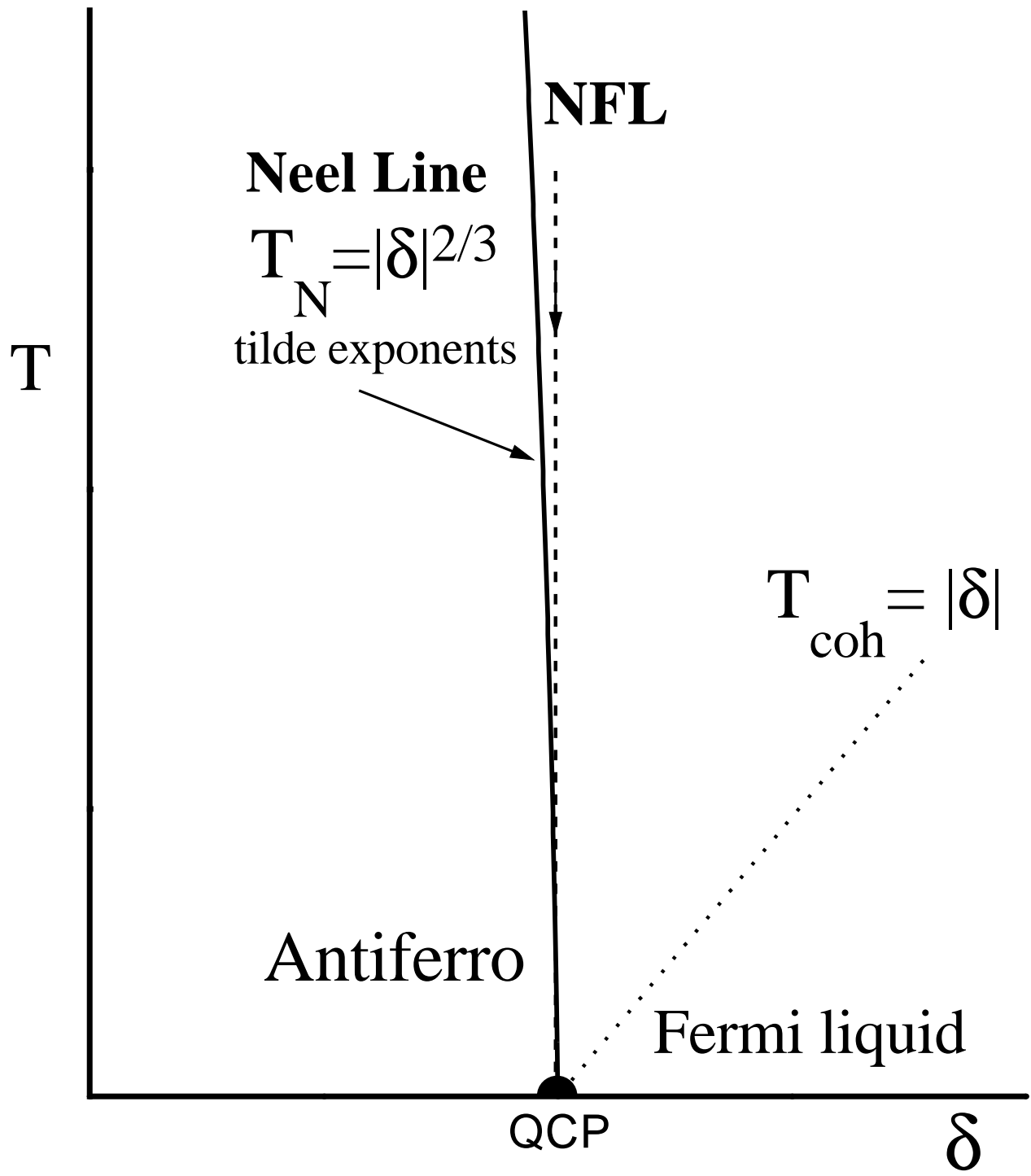
\end{document}